\documentclass[twocolumn,pra,amsmath,amssymb]{revtex4}

\usepackage{graphicx}
\usepackage{dcolumn}
\usepackage{bm}

\begin{document}

\preprint{APS/123-QED}

\title{Narrow Line Cooling and Momentum-Space Crystals}

\author{Thomas H. Loftus, Tetsuya Ido, Martin M. Boyd, Andrew D. Ludlow, and Jun Ye}
 \affiliation{JILA, National Institute of Standards and Technology and University of Colorado, Boulder, CO 80309-0440}

\date{\today}

\begin{abstract}
Narrow line laser cooling is advancing the frontier for
experiments ranging from studies of fundamental atomic physics to
high precision optical frequency standards. In this paper, we
present an extensive description of the systems and techniques
necessary to realize 689 nm $^1S_0$ - $^3P_1$ narrow line cooling
of atomic $^{88}$Sr. Narrow line cooling and trapping dynamics
are also studied in detail. By controlling the relative size of
the power broadened transition linewidth and the single-photon
recoil frequency shift, we show that it is possible to
continuously bridge the gap between semiclassical and quantum
mechanical cooling. Novel semiclassical cooling process, some of
which are intimately linked to gravity, are also explored.
Moreover, for laser frequencies tuned above the atomic resonance,
we demonstrate momentum-space crystals containing up to 26 well
defined lattice points. Gravitationally assisted cooling is also
achieved with blue-detuned light. Theoretically, we find the blue
detuned dynamics are universal to Doppler limited systems. This
paper offers the most comprehensive study of narrow line laser
cooling to date.
\end{abstract}

\maketitle

\section{Introduction}

Narrow line magneto-optical traps (MOTs) are rapidly becoming
powerful tools in a diverse array of experimental studies. These
unique and versatile systems have, for example, already been used
as integral components in obtaining fully maximized MOT
phase-space densities \cite{Katori3}, nuclear-spin based
sub-Doppler cooling \cite{Maru, JILA1d}, all-optical quantum
degenerate gases \cite{Takasu03}, and recoil-free spectroscopy of
both dipole allowed \cite{Katori4} and doubly-forbidden
\cite{Katori5} optical transitions. In the future, narrow line
MOTs promise to revolutionize the next generation of high
precision optical frequency standards \cite{Katori6, Curtis,
PTB}. Narrow line cooling, via the relative size of the
transition natural width $\Gamma$ and the single-photon recoil
frequency shift $\omega_R$, also displays a unique set of
thermal-mechanical laser cooling dynamics. In a previous paper
\cite{Loftus3}, we explored these behaviors by cooling $^{88}$Sr
on the $^1S_0$ - $^3P_1$ intercombination transition. In the
present work, we significantly expand upon this discussion and
provide a comprehensive description of the experimental
techniques used to realize and study $^1S_0$ - $^3P_1$ cooling
and trapping.

Fully understanding narrow line laser cooling requires first
clarifying the difference between broad and narrow Doppler
cooling lines. Broad lines, historically used for nearly every
laser cooling experiment \cite{Metcalf}, are defined by
$\Gamma$/$\omega_R$ $>>$ 1. The 461 nm $^{88}$Sr $^1S_0$ -
$^1P_1$ transition shown in Fig. 1(a), which typifies a broad
line, has, for example, $\Gamma$/$\omega_R$ $\sim$
3$\times$10$^3$. In this case, $\Gamma$, or more generally the
power-broadened linewidth $\Gamma_E$, is the natural energy
scale. Here, $\Gamma_E$ = $\Gamma\sqrt{1+s}$ is defined by the
saturation parameter $s$ = $I$/$I_S$ where $I$ ($I_S$) is the
single-beam peak intensity (transition saturation intensity).
Semiclassical physics thus governs the cooling process and the
photon recoil, although essential to quantitatively understanding
energy dissipation \cite{Wineland}, serves more as a useful
conceptual tool than a dominant player in system dynamics.
Moreover, gravity is essentially negligible since the ratio of
the maximum radiative force to the gravitational force, $R$ =
$\hbar k \Gamma$/2$mg$, is typically on the order of 10$^5$, where
$2\pi\hbar$, $k$, $m$, and $g$ are Plank's constant, the light
field wavevector, the atomic mass, and the gravitational
acceleration, respectively.

In contrast, narrow Doppler cooling lines are characterized by
$\Gamma$/$\omega_R$ $\sim$ 1. The 689 nm $^{88}$Sr $^1S_0$ -
$^3P_1$ transition used in this work, for example, has
$\Gamma$/$\omega_R$ = 1.6 where $\Gamma$/2$\pi$
($\omega_R$/2$\pi$ = $\hbar k^2$/4$\pi m$) is 7.5 kHz (4.7 kHz).
In this case, the relevant thermal-mechanical energy scale and
thus the underlying semiclassical or quantum mechanical nature of
the cooling depends on $s$. Details of a given cooling process are
then set by the laser detuning 2$\pi\delta$ = $\Delta$ =
$\omega_L - \omega_A$ where $\omega_L$ ($\omega_A$) is the laser
(atomic resonance) frequency. In particular, $\delta$ $<$ 0
$^1S_0$ - $^3P_1$ MOT dynamics can be divided into three
qualitatively distinct regimes, hereafter labeled (I - III),
defined by the relative size of $|\Delta|$, $\Gamma$ $\sim$
$\omega_R$, and $\Gamma_E$. In regime (III), corresponding to
trapping beam intensities on the order of $I_S$ = 3$\mu$W/cm$^2$
or $s$ $\sim$ 1, $\omega_R$ is the natural energy scale. Here,
single photon recoils and consequently, quantum physics, govern
trap dynamics. This situation enables limiting temperatures of
roughly half the photon recoil limit ($T_R$ =
2$\hbar\omega_R$/$k_B$ = 460 nK, where $k_B$ is Boltzmann's
constant) despite the incoherent excitation provided by the
trapping beams \cite{Metcalf, dalibard89}.

Conversely, for $s$ $\gg$ 1 the system evolves toward
semiclassical physics where $\Gamma_E$ $\gg$ $\omega_R$ and hence
$\Gamma_E$ is the dominant energy scale. In this case, cooling
and motional dynamics are determined by the relative size of
$\Gamma_E$ and $\Delta$. In regime (II), the $\Delta$ $<$
$\Gamma_E$ radiative force produces damped harmonic motion and,
in analogy to standard Doppler cooling, MOT thermodynamics set
entirely by the velocity dependence of the force \cite{Lett,
JILA1c}. For these conditions, the expected $\delta$- and
$s$-dependent temperature minima are observed, although with
values globally smaller than standard Doppler theory predictions.
Alternatively, in regime (I) where $\Delta$ $>$ $\Gamma_E$, the
atom-light interaction is dominated by single-beam photon
scattering and trap thermodynamics become intimately linked to
both the velocity and the {\it spatial} dependence of the force.
Here, gravity plays an essential role as the ratio $R$ for the
$^1S_0$ - $^3P_1$ transition is only $\sim$ 16. Consequently, the
atoms sag to vertical positions where the Zeeman shift balances
$\delta$, leading to $\delta$-independent equilibrium
temperatures.

Narrow line cooling also displays a unique set of $\delta$ $>$ 0
thermal and mechanical dynamics. For these experiments, the
atomic gas is first cooled to $\mu$K temperatures and then
$\delta$ is suddenly switched from $\delta$ $<$ 0 to $\delta$ $>$
0. Subsequently, the sample evolves from a thermal distribution
to a discrete set of momentum-space packets whose alignment
matches lattice points on a three-dimensional (3D)
face-centered-cubic crystal \cite{Com1}. Up to 26 independent
packets are created with $\delta$- and $s$-dependent lattice
point filling factors. Note this surprising behavior occurs in
the setting of incoherent excitation of a non-degenerate thermal
cloud. To obtain qualitative insight into the basic physics, we
begin with an analytic solution to the one-dimensional (1D)
semiclassical radiative force equation. Here, we show that
$\delta$ $>$ 0 excitation enables "positive feedback"
acceleration that efficiently bunches the atoms into discrete
sets of $\delta$- and $s$-dependent velocity space groups. A
simple generalization of the 1D model is then used to motivate
the experimentally observed 3D lattice structure. This intuitive
picture is then confirmed with numerical calculations of the
final atomic velocity and spatial distributions. Using the
numerical calculations, we also show that $\delta$ $>$ 0
momentum-space crystals are a universal feature of standard
Doppler cooling and that observations should be possible,
although increasingly impractical, with broad line molasses.

Finally, we demonstrate that $R$ directly influences $\delta$ $>$
0 thermodynamics, enabling cooling around a $\delta$- and
$s$-dependent velocity $v_0$ where gravity balances the radiative
force. Observed values for $v_0$ agree well with numerical
predictions while cooling is evident in distinctly asymmetric
cloud spatial distributions that appear in both numerical
calculations of the cooling process and the experiment. As with
momentum crystal formation, gravitationally assisted $\delta$ $>$
0 cooling is universal to Doppler limited systems. In the more
typical case where $R$ $\sim$ 10$^5$, however, equilibrium
temperatures ($v_0$) are on the order of hundreds of milli-Kelvin
(100 m/s) rather than the more useful micro-Kelvin ($\sim$ 10
cm/s) values achieved with narrow lines.

The remainder of this paper is organized as follows. Section II
gives an overview of the 461 nm $^1S_0$ - $^1P_1$ MOT used to
pre-cool $^{88}$Sr to $\sim$ 2.5 mK. $^3P_2$ metastable
excited-state magnetic traps that are continuously loaded by the
$^1S_0$ - $^1P_1$ cooling cycle are also described. Section III
details the highly stabilized 689 nm light source and the process
used to transfer $^{88}$Sr from the $^1S_0$ - $^1P_1$ MOT to the
$^1S_0$ - $^3P_1$ MOT. Descriptions of the techniques used to
control and study $^1S_0$ - $^3P_1$ laser cooling are also
provided. Sections IV and V then focus on $\delta$ $<$ 0
mechanical and thermal dynamics, respectively. Finally, section
VI explores $\delta$ $>$ 0 cooling and momentum-space crystals.
Conclusions are given in section VII.

\section{$^1S_0$ - $^1P_1$ MOT Pre-Cooling}

\begin{figure}
\resizebox{8.2cm}{!}{
\includegraphics{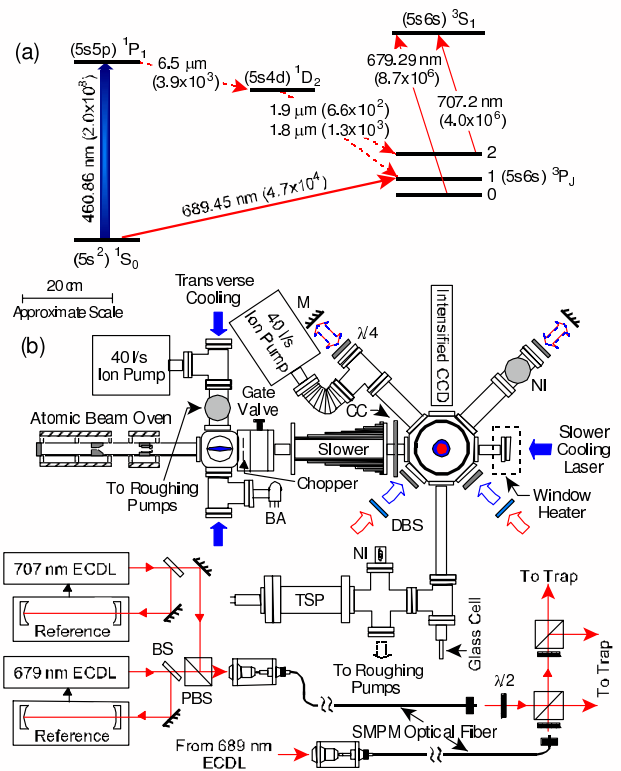}}
\caption{\label{Fig1} (a) Partial $^{88}$Sr Energy level diagram.
Numbers in parentheses give Einstein A coefficients (in
s$^{-1}$). Wavelengths are in vacuum. (b) Top (x-y plane) view of
the Sr cooling and trapping apparatus. Blue (red) arrows
represent 461 nm (689 nm, 679 nm, and 707 nm) trapping (trapping
and re-pumping) beams. M, mirror; $\lambda$/4, dual wavelength
quarter-wave plate; $\lambda$/2, half-wave plate; DBS, dichroic
beamsplitter; PBS, polarization beamsplitter; BS, beamsplitter;
SMPM, single-mode polarization maintaining; ECDL, external-cavity
diode laser; BA, Bayard-Alpert vacuum gauge; NI, nude vacuum
gauge, TSP, Titanium sublimation pump; CC, compensation coil.}
\end{figure}

Pre-cooling $^{88}$Sr to milli-Kelvin temperatures is an
essential requirement for observing $^1S_0$ - $^3P_1$ cooling and
trapping dynamics \cite{Katori3}. For this purpose, as shown by
Fig. 1(b), we use a standard six-beam 461 nm $^1S_0$ - $^1P_1$
MOT that is loaded by a Zeeman slowed and transversely cooled
atomic beam. The atomic beam is generated by an effusion oven (2
mm nozzle diameter) whose output is angularly filtered by a 3.6
mm diameter aperture located 19.4 cm from the oven nozzle.
Separate heaters maintain the oven body (nozzle) at 525 $^o$C
(725 $^o$C), resulting in a measured flux (divergence half-angle)
of 3$\times$10$^{11}$ atoms/s (19 mrad). The atomic beam is then
transversely cooled by 2-dimensional 461 nm optical molasses. The
elliptical cross-section molasses laser beams have a 1/e$^2$
diameter of $\sim$ 3 cm ($\sim$ 4 mm) along (normal to) the
atomic beam propagation axis, contain 10 - 20 mW of power, and
are detuned from the $^1S_0$ - $^1P_1$ resonance by -15 MHz.
Stray magnetic fields in the transverse cooling region are less
than 1 G. Subsequently, the atomic beam passes through a 6.4 mm
diameter electro-mechanical shutter and a gate valve that allows
the oven to be isolated from the rest of the vacuum system.

After exiting the transverse cooling region, the atomic beam
enters a water cooled 20 cm long constant deceleration $\sigma^-$
Zeeman slower \cite{slower} with a peak magnetic field of $\sim$
600 G, corresponding to a capture velocity of $\sim$ 500 m/s. The
461 nm Zeeman slower cooling laser is detuned from the $^1S_0$ -
$^1P_1$ resonance by -1030 MHz, contains 60 mW of power, and is
focused to approximately match the atomic beam divergence. The
window opposite the atomic beam is a z-cut Sapphire optical flat
that is vacuum sealed via the Kasevich technique \cite{Kasevich},
broadband anti-reflection coated on the side opposite the
chamber, and heated to 200 $^o$C to prevent the formation of Sr
coatings. The alternating current window heater, which produces a
small stray magnetic field, is only operated during the $^1S_0$ -
$^1P_1$ MOT pre-cooling phase. A separate compensation coil
reduces the slower magnetic field magnitude (gradient) to $<$ 100
mG ($<$ 8 mG/cm) at the trapping region, located 15 cm from the
slower exit. Stray magnetic fields at the trap are further nulled
by three sets of orthogonally oriented Helmholtz pairs.

The trapping chamber is a cylindrical octagon with six 2 3/4''
(two 6'') ports in the horizontal x-y plane (along gravity
z-axis). All windows are broadband anti-reflection coated. The
MOT anti-Helmholtz coils, oriented such that the axial magnetic
field gradient dB$_z$ lies along gravity, are mounted on a
computer controlled precision linear track that allows the coil
center to be translated from the trapping chamber to a 1 cm
$\times$ 1 cm $\times$ 4 cm rectangular glass cell. The coils are
constructed to provide window-limited optical access to the
geometric center of the trapping chamber and produce axial
gradients of 0.819 G/(cm-A). The gradient is linear over a
spatial range of $\sim$ $\pm$ 3 cm in both the axial and
transverse directions. Current in the coils is regulated by a
computer controlled servo and monitored with a Hall probe. For
the $^1S_0$ - $^1P_1$ MOT, the axial magnetic field gradient
dB$_z$/dz = dB$_z$ = 50 G/cm.

The trapping chamber is evacuated by a 40 $l$/s Ion pump and a
Titanium sublimation pump while the oven chamber uses a 40 $l$/s
Ion pump. The two chambers are separated by a 6.4 mm $\times$ 45
mm cylindrical differential pumping tube located between the
electro-mechanical shutter and the gate valve. Typical vacuum
levels in the oven, trapping, and glass cell chambers during
operation of the atomic beam are 2$\times$10$^{-8}$ Torr,
1.5$\times$10$^{-9}$ Torr, and 3$\times$10$^{-10}$ Torr,
respectively.

461 nm cooling and trapping light is produced by frequency
doubling the output from a Ti:Sapphire laser in two external
buildup cavities \cite{doubler} that together produce $>$ 220 mW
of single-mode light. The 461 nm light is then offset locked to
the $^1S_0$ - $^1P_1$ resonance by saturated absorption feedback
to the Ti:Sapphire laser. Relative frequencies of the transverse
cooling, Zeeman slower, and trapping laser beams are controlled
with acousto-optic modulators (AOMs) which are also used as
shutters. Additional extinction of 461 nm light is provided by
electro-mechanical shutters. The intensity stabilized trapping
beams have 1/e$^2$ diameters of $\sim$ 3 cm, are detuned from the
$^1S_0$ - $^1P_1$ resonance by -40 MHz, and typically have a
total power of 30 mW. For these settings, we find the $^1S_0$ -
$^1P_1$ MOT population is maximized. Further increases in, for
example, the trapping beam power simply increases the cloud
temperature. $^1S_0$ - $^1P_1$ MOTs are monitored with a
charge-coupled-device (CCD) camera and a calibrated photodiode.
Typical trap lifetimes, populations, and temperatures are 20 ms,
3$\times$10$^7$, and $\sim$ 2.5 mK, respectively.

\begin{figure}
\resizebox{8.4cm}{!}{
\includegraphics{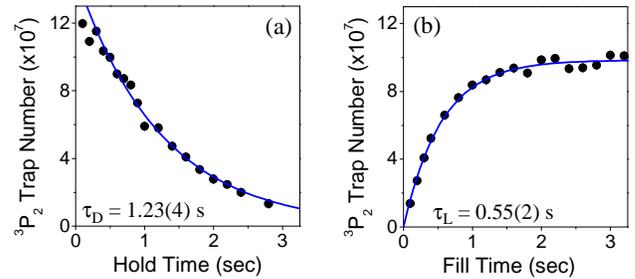}}
\caption{\label{Fig2} $^3P_2$ Magnetic trap population versus (a)
hold time for a fixed 2 s loading time and (b) loading time at a
hold time of 400 ms. Solid lines are exponential fits. $\tau_D$
($\tau_L$) is the measured exponential decay (loading) time.}
\end{figure}

As shown by Fig. 1(a), operation of the $^1S_0$ - $^1P_1$ MOT
efficiently populates the ground-state-like $^3P_2$ metastable
excited-state ($\sim$ 500 s radiative lifetime \cite{Katori1})
via $^1P_1$ $\rightarrow$ $^1D_2$ $\rightarrow$ $^3P_2$ radiative
decay. Consequently, $^1S_0$ - $^1P_1$ MOT lifetimes are typically
limited to 10 - 50 ms \cite{Loftus4, JILA1a, JILA1b}. To overcome
this loss process, $^3P_2$ population is re-pumped to the $^1S_0$
ground-state via the $^3P_2$ $\rightarrow$ $^3S_1$ $\rightarrow$
$^3P_1$ $\rightarrow$ $^1S_0$ channel by driving the 707 nm
$^3P_2$ - $^3S_1$ and 679 nm $^3P_0$ - $^3S_1$ transitions with
two external cavity diode lasers (ECDLs). Each laser is locked to
a reference cavity that is simultaneously locked to a frequency
stabilized helium neon laser. Double-passed AOMs are then used to
tune the absolute laser frequencies. After passing through a
single-mode polarization-maintaining optical fiber, the
co-propagating 707 nm and 679 nm laser beams are expanded to a
1/e$^2$ diameter of $\sim$ 1 cm and delivered to the $^1S_0$ -
$^1P_1$ MOT. An AOM located before the beam expansion optics
allows for rapidly turning the beams either on or off ($<$ 1
$\mu$s transition time). At the trap, the 707 nm (679 nm) beam
contains 1.5 mW (2.5 mW) of power, resulting in an optical
re-pumping time of $<$ 100 $\mu$s. With both lasers operating,
the $^1S_0$ - $^1P_1$ MOT population and lifetime are typically
enhanced by 10$\times$ and 15$\times$, respectively, with the
former value limited by atomic beam induced trap loss.

Along with contributing to $^1S_0$ - $^1P_1$ MOT loss, $^1P_1$
$\rightarrow$ $^1D_2$ $\rightarrow$ $^3P_2$ radiative decay
continuously loads milli-Kelvin atoms in the $^3P_2$(m=1,2)
states into a magnetic trap formed by the $^1S_0$ - $^1P_1$ MOT
quadrupole magnetic field \cite{Loftus1, Katori2, Loftus2, JILA1b,
Killian, Hemmerich}. Importantly, these samples are expected to
display a wealth of binary collision resonances that arise due to
an interplay between anisotropic quadrupole interactions and the
local magnetic field \cite{Derevianko, GreenePRL, GreenePRA}.
Qualitatively similar processes and hence, collision resonances,
are predicted for polar molecules immersed in electrostatic fields
\cite{Bohn}. Studies of metastable Sr collision dynamics will thus
likely impact the understanding of a diverse range of physical
systems. To pursue these studies, we plan to first load $^3P_2$
state atoms into the quadrupole magnetic trap and then
mechanically translate \cite{Lewandowski} the sample to the glass
cell chamber. Subsequently, a tight Ioffe-Pritchard magnetic trap
will be used to perform a variety of collision experiments.

\begin{figure}
\resizebox{8.2cm}{!}{
\includegraphics{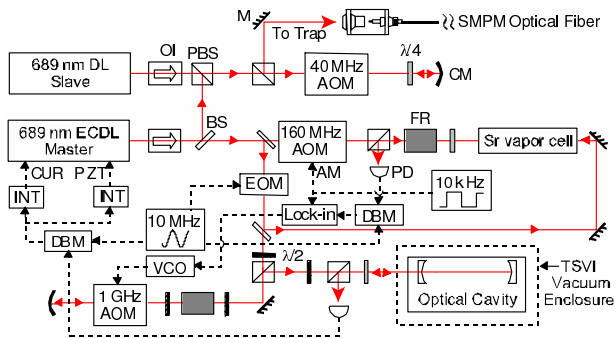}}
\caption{\label{Fig3} 689 nm laser system. Solid (dashed) lines
indicate optical beams (electrical connections). $\lambda$/4,
quarter-wave plate; $\lambda$/2, half-wave plate; M, mirror; CM,
curved mirror; PBS, polarization beamsplitter; BS, beamsplitter;
OI, optical isolator; SMPM, single-mode polarization maintaining;
ECDL, external-cavity diode laser; DL, diode laser; TSVI,
temperature stabilized and vibration isolated; PD, photodiode;
AOM, acousto-optic modulator; EOM, electro-optic modulator; DBM,
double balanced mixer; INT, integrator; AM, amplitude modulation
input; CUR (PZT), current (piezo-electric mounted grating)
modulation input.}
\end{figure}

As a first step in this direction we have loaded $>$ 10$^8$
$^3P_2$(m=1,2) state atoms into the quadrupole magnetic trap and
achieved trap lifetimes $>$ 1 s. For these measurements, the
$^3P_2$ trap is first loaded by operating the $^1S_0$ - $^1P_1$
MOT for a variable fill time. The 461 nm cooling and trapping
lasers and the atomic beam are then switched off. Following a
variable hold time, the 461 nm trapping beams and the 707 nm and
679 nm re-pumping beams are switched on, enabling the $^3P_2$
magnetic trap population N$_M$ to be determined from 461 nm
fluorescence \cite{Katori2}. Figure 2(a) shows N$_M$ versus hold
time for a fill time of 2 s while Fig. 2(b) gives N$_M$ versus
fill time at a fixed hold time of 400 ms. For both, dB$_z$ = 50
G/cm, giving a magnetic trap depth of $\sim$ 30 mK. The $^1S_0$ -
$^1P_1$ MOT lifetime (steady-state population) is 16 ms
(2.8$\times$10$^7$). The observed 1.23(4) s exponential decay
time agrees well with the expected $\sim$ 1.3 s vacuum limited
value extracted from measurements reported in Ref. \cite{Killian}
at higher pressures. In contrast, the significantly shorter
0.55(2) s fill time implies that additional loss processes are
operative during the loading phase, possibly due to interactions
between atoms in the $^1D_2$ or $^3P_2$ states with $^1S_0$ -
$^1P_1$ MOT atoms or the atomic beam \cite{Stuhler}. This idea is
supported by the magnetic trap loading rate. For the parameters
used here, the observed rate of 2.7(5)$\times$10$^8$ atoms/s is
$\sim$ 2$\times$ smaller than the theoretically predicted value
of 5.4(1.8)$\times$10$^8$ atoms/s \cite{Loftus1} where the theory
only accounts for relevant branching ratios in the $^1P_1$
$\rightarrow$ $^1D_2$ $\rightarrow$ $^3P_2$(m=1,2) radiative
cascade. Similar discrepancies between observed and predicted
loading rates were reported in Ref. \cite{Killian}.

\section{$^1S_0$ - $^3P_1$ MOT Loading and Detection}

Exploring 689 nm $^1S_0$ - $^3P_1$ narrow line cooling dynamics
requires a laser system whose short-term linewidth is small
compared to the 7.5 kHz transition natural width. In addition,
the absolute laser frequency must be referenced, with similar
stability, to the $^1S_0$ - $^3P_1$ transition and be tunable
over a range of $\sim$ 10 MHz. Figure 3 shows the 689 nm laser
stabilization and control system consisting of a master-slave
ECDL, a temperature stabilized and vibration isolated passive
optical reference cavity, and a Sr saturated absorption
spectrometer.

The linewidth of the master ECDL is first narrowed by locking the
laser to a stable optical reference cavity via the
Pound-Drever-Hall technique \cite{PDH}. The cavity consists of
high reflectivity Zerodour substrate mirrors that are optically
contacted to a Zerodour spacer. The measured cavity finesse (free
spectral range) at 689 nm is $\sim$ 3800 (488.9 MHz), giving a
linewidth for the TEM$_{00}$ mode of $\sim$ 130 kHz. To isolate
the cavity from environmental perturbations, the cavity is
suspended by two thin wires inside a temperature stabilized can
that is evacuated to $<$ 10$^{-6}$ Torr and mounted on vibration
damping material. The absolute cavity frequency is tuned by
double-passing the 689 nm light sent to the cavity through a 1
GHz AOM. The electronic feedback, with an overall bandwidth of
$\sim$ 2 MHz is divided into a slow loop that adjusts the
piezo-electric mounted ECDL grating and a fast loop that couples
to the diode laser current. With the cavity lock engaged, in-loop
analysis shows that jitter in the cavity-laser lock is $<$ 1 Hz.

To further evaluate the performance of the 689 nm laser system,
the short-term laser linewidth is determined by beating the
cavity-locked 689 nm light against a femto-second comb that is
locked to a second optical cavity \cite{Jones}. Here, we find a
short-term linewidth of $<$ 300 Hz, where the measurement is
limited by the optical fiber connecting the 689 nm light to the
femto-second comb. Next, the absolute stability of the 689 nm
laser is evaluated by beating the cavity locked 689 nm light
against a femto-second comb that is locked to a Hydrogen maser via
a fiber optical link to NIST \cite{maser}. From these
measurements, the laser-cavity system drifts $\sim$ 400 mHz/s and
has a 1 s stability of $<$ 4$\times$10$^{-13}$ (i.e., $<$ 180 Hz)
with the former value limited by cavity drift and the latter
value limited by the effective maser noise floor. To eliminate
the slow 689 nm cavity induced drift, the master ECDL is next
locked to the $^1S_0$ - $^3P_1$ resonance via saturated
absorption feedback to the 1 GHz AOM. For added stability, a DC
magnetic field is applied to the Sr vapor cell and the
spectrometer is set to perform frequency modulation spectroscopy
on the $^1S_0$ - $^3P_1$(m=0) transition. With the system fully
locked, the 1 s stability (drift rate) is then $\sim$
4$\times$10$^{-13}$ ($<$ 80 mHz/s).

A portion of the 689 nm master ECDL output is next used to
injection lock a 689 nm slave diode laser. The slave laser
output, after double-passing through an AOM used for frequency
shifting and intensity chopping, is then coupled into a
single-mode polarization-maintaining optical fiber. Upon exiting
the fiber, the 689 nm light, containing up to 6 mW of power, is
expanded to a 1/e$^2$ diameter of 5.2 mm and divided into three
equal intensity trapping beams. Dichroic beamsplitters are then
used to co-align the 689 nm and 461 nm trapping beams. The
trapping beam waveplates are 3$\lambda$/4 at 461 nm and
$\lambda$/4 at 689 nm.

\begin{figure}
\resizebox{8.5cm}{!}{
\includegraphics{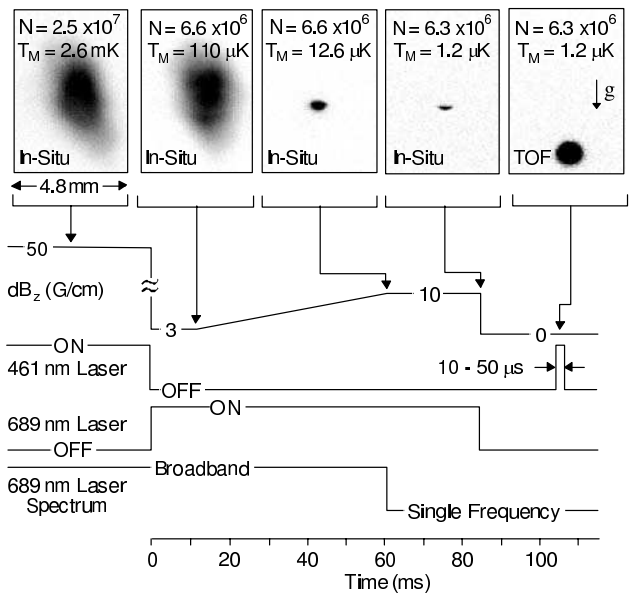}}
\caption{\label{Fig4} $^1S_0$ - $^3P_1$ MOT timing diagram.
Images at top show the atomic cloud at various stages of the
cooling and compression process. From left to right, the first
four images show in-situ images of the cloud while the last frame
shows the cloud after 25 ms of free expansion from the $\delta$ =
-520 kHz, s = 75 single-frequency MOT. N, cloud population;
T$_M$, cloud temperature.}
\end{figure}

$^1S_0$ - $^3P_1$ cooling and trapping dynamics are monitored
either by in-situ or time-of-flight (TOF) fluorescence images
collected with an intensified CCD camera. The camera is set to
view the cloud in either the horizontal x-y plane or nearly along
gravity. The x-y plane (along gravity) images have a spatial
resolution of 21 $\mu$m/pixel (37 $\mu$m/pixel). For in-situ
images, the 461 nm trapping beams are pulsed on for 10 - 50 $\mu$s
immediately after the atoms are released from the trap while for
TOF images, the atoms are allowed to first freely expand for a
variable amount of time. We have verified that in-situ images
recorded with 461 nm pulses are identical to direct images of the
in-trap 689 nm fluorescence aside from an improved
signal-to-noise ratio. Typical TOF flight times are 20 - 35 ms.
To determine cloud temperatures, gaussian fits are performed to
both the in-situ and TOF images. The temperature T$_M$ is then
given by T$_M$ = ($m$/4$k_B$t$_F$$^2$)(R$_F^2$ - r$^2$)$^2$ where
t$_F$ is the flight time and R$_F$ (r) is the TOF (in-situ)
1/e$^2$ radius of the cloud.

Figure 4 depicts the $^1S_0$ - $^3P_1$ MOT loading procedure
\cite{Katori3}. As outlined above, $\sim$ 3$\times$10$^7$ atoms
are first pre-cooled to $\sim$ 2.5 mK in a $^1S_0$ - $^1P_1$ MOT.
(Note that for the remainder of this paper, the $^1S_0$ - $^1P_1$
MOT is loaded without the 707 nm and 679 nm re-pumping lasers.) At
time t = 0, the 461 nm light and the atomic beam shutter are
switched off, dB$_z$ is rapidly lowered to 3 G/cm, and
red-detuned, broadband frequency modulated 689 nm trapping beams
are turned on. 10 ms later and for the following 50 ms, the cloud
is compressed by linearly increasing dB$_z$ to 10 G/cm. Frequency
modulation parameters for the 689 nm trapping beams are set to
provide complete spectral coverage of the $^1S_0$ - $^1P_1$ MOT
Doppler profile and, as shown below, manipulate the cloud size at
the end of the magnetic field ramp. Subsequently, at t = 60 ms,
the frequency modulation is turned off and the atoms are held in
a single-frequency MOT. Overall, as shown by the images at the
top of Fig. 4, this process reduces the sample temperature by
more than three orders of magnitude while only reducing the cloud
population by a typical factor of 3 - 4, giving final
temperatures of $\sim$ 1 $\mu$K and populations of $\sim$ 10$^7$.
Typical single frequency trap lifetimes and spatial densities are
$\sim$ 1 s and $\sim$ 5$\times$10$^{11}$ cm$^{-3}$, respectively.

\section{$\delta$ $<$ 0 Mechanical Dynamics}

As outlined in section I, $\delta$ $<$ 0 $^1S_0$ - $^3P_1$ MOTs
provide a unique opportunity to explore three qualitatively
distinct laser cooling regimes whose underlying mechanics are
governed by either semiclassical or quantum mechanical physics.
In this and the following section, we will describe the unique
experimental signatures for regimes (I) - (III) and give detailed
explanations for the observed trapped atom behavior.

\begin{figure}
\resizebox{7.8cm}{!}{
\includegraphics{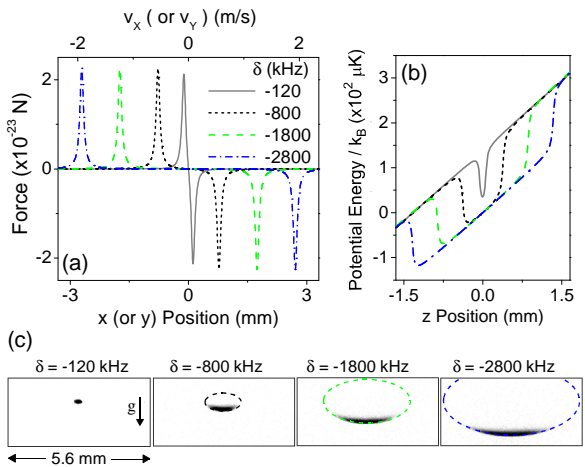}}
\caption{\label{Fig5} (a) Semiclassical radiative force versus
position (bottom axis, $v_x$ = $v_y$ = 0) and velocity (upper
axis, x = y = 0) for a range of detunings. Corresponding (b) trap
potential energy in the z direction and (c) in-situ images of the
$^1S_0$ - $^3P_1$ MOT. Dashed lines in (c) are calculated maximum
force contours. For each, $s$ = 248.}
\end{figure}

Insight into regime (I) and (II) thermal-mechanical dynamics is
provided by the semiclassical radiative force equation \cite{Lett}
\begin{eqnarray}
F(\vec{v}, \vec{x}) &=& \frac{\hbar \vec{k}\Gamma}{2}[\frac{s}{1
+ s' + 4(\Delta -
\vec{k}\cdot\vec{v} - \mu\vec{dB}\cdot\vec{x})^2/\Gamma^2}\nonumber \\
&-& \frac{s}{1 + s' + 4(\Delta + \vec{k}\cdot\vec{v} + \mu
\vec{dB}\cdot\vec{x})^2/\Gamma^2}]\nonumber \\ &-& m\vec{g}.
\label{SCF}
\end{eqnarray}
where $\vec{x}$ = $\{x,y,z\}$, $\vec{dB}$ = $\{dB_x,dB_y,dB_z\}$,
and $\mu$ = ($g_J\mu_B$/$\hbar$) where $g_J$ = 1.5 ($\mu_B$) is
the $^3P_1$ state Lande g-factor (Bohr magneton). $s'$ $>$ $s$
accounts, along a single axis, for saturation induced by the
remaining four trapping beams. Figure 5(a) shows the $^1S_0$ -
$^3P_1$ radiative force at $s$ = $s'$ = 248 for a range of
$\delta$ values while Fig. 6(a) shows the force at $\delta$ =
-520 kHz for a range of $s$ = $s'$ values. As described in
section I, the qualitative nature of the force and hence the
resulting trap mechanical dynamics depends on the relative size of
$\Delta$ and $\Gamma_E$. For regime (I), corresponding to
$|\delta|$ $>$ 120 kHz in Fig. 5(a) or $s$ $<$ 248 in Fig. 6(a),
$\Delta$ $>$ $\Gamma_E$ and the 3D radiative force acts only along
a thin shell volume marking the outer trap boundary. Here, the
trap boundary roughly corresponds to positions where the radiative
force is peaked. This situation, as shown by Figs. 5(b) and 6(b),
produces a box potential with a gravitationally induced z-axis
tilt. Hence, in the x-y plane, motion consists of free-flight
between hard wall boundaries while along the z-axis, mechanical
dynamics are set by the relative size of the radiative force
``kicks," gravity, and the cloud thermal energy. As shown in
section V, the thermal energy is small compared to the
gravitational potential energy. Moreover, the ratio of the
maximum radiative force to the gravitational force, $R$ $\sim$
16. Thus, the atoms sink to the bottom of the trap where they
interact, along the z-axis, with only the upward propagating
trapping beam.

\begin{figure}[t]
\resizebox{7.8cm}{!}{
\includegraphics{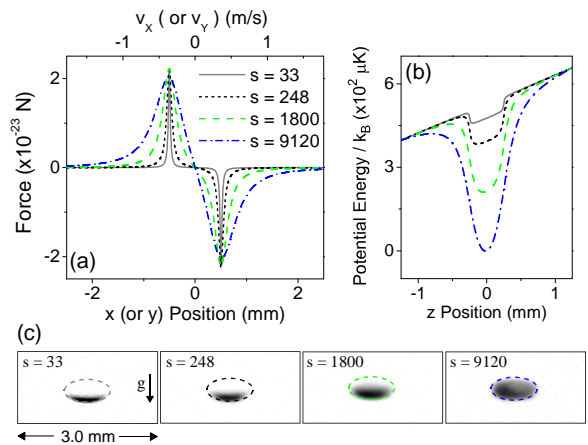}}
\caption{\label{Fig6} (a) Semiclassical radiative force versus
position (bottom axis, $v_x$ = $v_y$ = 0) and velocity (upper
axis, x = y = 0) for a range of intensities. Corresponding (b)
trap potential energy in the z direction and (c) in-situ images
of the $^1S_0$ - $^3P_1$ MOT. Dashed lines in (c) are calculated
maximum force contours. For each, $\delta$ = -520 kHz.}
\end{figure}

As $\delta$ decreases in Fig. 5(a) or $s$ increases in Fig. 6(a),
the trap mechanically evolves to regime (II) where $\Delta$ $<$
$\Gamma_E$ produces a linear restoring force and hence, damped
harmonic motion \cite{Lett, JILA1c}. Consequently the trap
potential energy assumes the U-shaped form familiar from standard
broad line Doppler cooling. As the trap moves more fully into
regime (II), perturbations to the potential energy due to gravity
become less pronounced. One expects, therefore, that the cloud
aspect ratio will evolve toward the 2:1 value set by the
quadrupole magnetic field.

\begin{figure}[b]
\resizebox{7cm}{!}{
\includegraphics{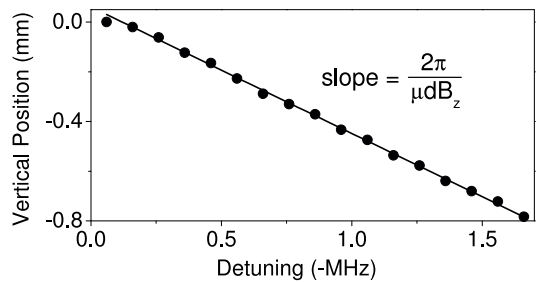}}
\caption{\label{Fig7} Vertical cloud position versus $\delta$ for
$s$ = 248 and dB$_z$ = 10 G/cm. The solid line is a linear fit.}
\end{figure}

The intuitive descriptions developed above are directly confirmed
by Figs. 5(c) and 6(c) which show in-situ images of the $^1S_0$ -
$^3P_1$ MOT along with overlaid maximum force contours calculated
from Eq. \ref{SCF}. For excitation conditions corresponding to
regime (II), the cloud approaches the 2:1 quadrupole magnetic
field aspect ratio. In contrast, for regime (I) the cloud x-y
width is determined largely by the separation between x-y force
maxima or alternatively, by the wall separation for the x-y
potential energy box. In the vertical direction, the atoms sink
to the bottom of the trap where the lower cloud boundary $z_0$ is
defined by the location of the z-axis potential energy minima
which is, in turn, proportional to the position where the Zeeman
shift matches the laser detuning. As $\delta$ increases, $z_0$
shifts vertically downward, an effect predicted in Fig. 5(b) and
clearly revealed in Fig. 5(c). To quantify this relationship,
Fig. 7 shows $z_0$ versus $\delta$ along with a linear fit to the
data giving dz$_0$/d$\delta$ = 2$\pi$/($\mu$dB$_z$) = 0.509(4)
$\mu$m/kHz, in agreement at the 5$\%$ level with the expected
linear slope of 0.478(2) $\mu$m/kHz.

\begin{figure}[t]
\resizebox{7.5cm}{!}{
\includegraphics{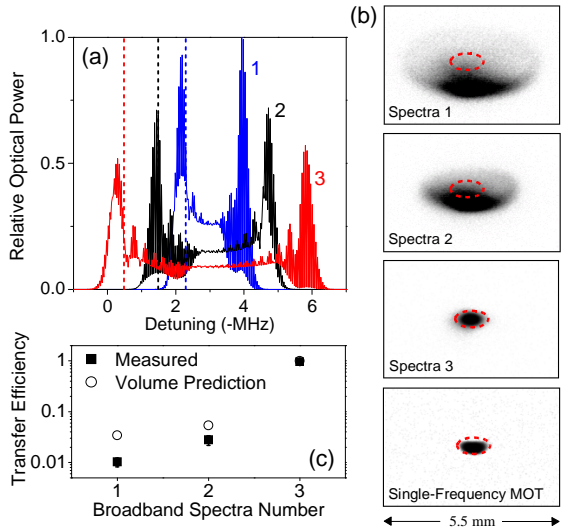}}
\caption{\label{Fig8} Modematching broadband cooled atoms to the
single-frequency MOT. (a) Broadband modulation spectra. Dashed
vertical lines give the effective detuning corresponding to the
measured cloud size. (b) In-situ images of the broadband cooled
atoms. For each, the temperature is 8.5(0.8) $\mu$K. Dashed lines
give the maximum force contour for the $\delta$ = -520 kHz, $s$ =
75 single-frequency MOT shown in the bottom frame. (c) Broadband
to single-frequency MOT transfer efficiency versus broadband
spectra number.}
\end{figure}

Unique $^1S_0$ - $^3P_1$ MOT mechanical dynamics are also
manifest in the transfer between the broadband and
single-frequency cooling stages. Here, it is important to note
that following broadband cooling, the cloud vertical and
horizontal 1/e$^2$ radii (r$_z$ and r$_h$, respectively) are set
by the final value for dB$_z$ and the spectral separation between
the blue edge of the modulation spectrum and the $^1S_0$ -
$^3P_1$ resonance. Figure 8(a) shows three typical modulation
spectra while Fig. 8(b) shows in-situ images of the corresponding
broadband cooled clouds at the end of the magnetic field ramp.
For each, the measured temperature is 8.5(0.8) $\mu$K. Overlaid
dashed lines give  the maximum force contour for the subsequent
$\delta$ = -520 kHz, $s$ = 75 single-frequency MOT shown in the
bottom frame. Here, dynamics similar to Fig. 5 are observed: as
the broadband modulation moves closer to resonance, the cloud
density distribution becomes more symmetric and the aspect ratio
evolves toward 2:1. Figure 8(c) shows the broadband to
single-frequency MOT transfer efficiency TE versus broadband
spectra number. Clearly, TE increases as the overlap between the
broadband cooled cloud and the single-frequency MOT force contour
increases, indicating that TE can be optimized by ``mode-matching"
r$_z$ and r$_h$ to the single-frequency MOT box. This idea is
supported by open circles in Fig. 8(c) which give predicted TE
values based on the relative volume of the broadband cooled cloud
and the single-frequency radiative force curve. Overall, the
prediction reproduces observed TE values with a discrepancy for
spectra 1 and 2 that likely arises from the broadband MOT density
distribution.

\begin{figure}[t]
\resizebox{7cm}{!}{
\includegraphics{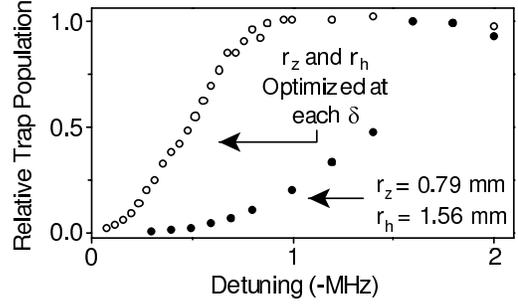}}
\caption{\label{Fig9} Relative trap population versus $\delta$
when r$_z$ and r$_h$ are optimized at each point (open circles) or
fixed at r$_z$ = 0.79 mm and r$_h$ = 1.56 mm (solid circles). For
both, s = 248 and dB$_z$ = 10 G/cm.}
\end{figure}

Further evidence for broadband to single-frequency mode-matching
is provided by the single-frequency MOT population N versus
$\delta$ shown in Fig. 9. Here, solid circles give N versus
$\delta$ when r$_z$ = 0.79 mm and r$_h$ = 1.56 mm. According to
the Fig. 5 model, the single-frequency MOT acquires these
dimensions at $\delta$ = -1637 kHz, which lies within 2$\%$ of
the $\delta$ value where N is peaked. Hence, we again find that
optimal transfer occurs when r$_z$ and r$_h$ are matched to the
single-frequency radiative force contour. Additional confirmation
for this effect is provided by open circles in the figure which
give N when r$_z$ and r$_h$ are optimized at each $\delta$. To
understand the trend in this case toward smaller N as $\delta$
decreases, note the non-zero slope for the blue edge of the
modulation spectrum sets the minimum effective detuning
$\delta_m$ that can be used during the broadband cooling phase.
In analogy to Fig. 5, $\delta_m$ sets minimum values for r$_z$
and r$_h$. Thus mode-matching becomes progressively more
difficult as $\delta$ decreases and the single-frequency
radiative force contour shrinks, leading to the observed decrease
in N.

\section{$\delta$ $<$ 0 Thermodynamics}

$\delta$ $<$ 0 $^1S_0$ - $^3P_1$ MOTs display a rich variety of
thermodynamic behaviors that are directly linked to the
mechanical dynamics explored in section IV. Figure 10(a) shows
the MOT equilibrium temperature T$_M$ versus $\delta$ for
saturation parameters ranging from $s$ = 7.5 to $s$ = 1800. For
large $\delta$ and $s$, corresponding to regime (I), $\Delta$ $>$
$\Gamma_E$ $\gg$ $\Gamma$, T$_M$ is basically
$\delta$-independent. Insight into this behavior is provided by
Fig. 11(a), which details the unique regime (I) connection
between trap thermodynamics, the spatial dependence of the
radiative force, and the relative size of the radiative force and
gravity. Recall that for $\Delta$ $>$ $\Gamma_E$, the cloud sags
to the bottom of the trap where interactions occur, along the
vertical z-axis, with only the upward propagating trapping beam.
Moreover, due to polarization considerations and the free-flight
motion executed by atoms in the horizontal plane, horizontal beam
absorption rates are more than 4$\times$ smaller than the
vertical rate. Trap thermodynamics, therefore, are dominated by a
balance between gravity and the radiative force due to the upward
propagating beam. Here, it is important to realize that as
$\delta$ changes, the z-axis atomic position $z_0$ self adjusts
such that the effective detuning, $\Delta$ - $\mu$dB$_z z_0$,
remains constant. Consequently the trap damping and diffusion
coefficients, and thus the equilibrium temperature, remain
constant.

\begin{figure}[t]
\resizebox{7.8cm}{!}{
\includegraphics{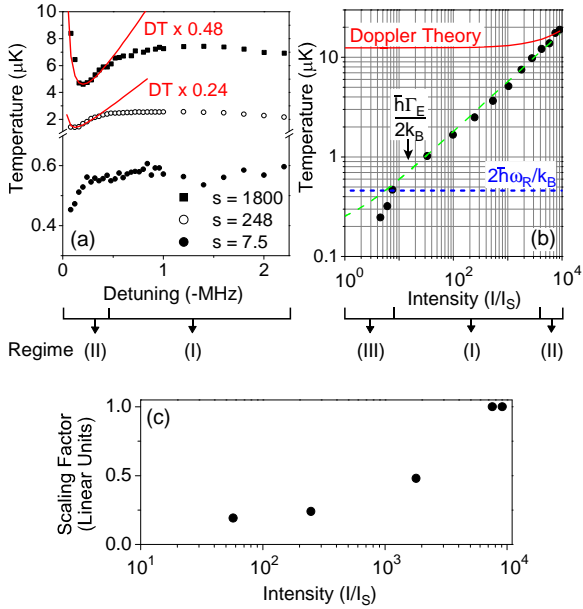}}
\caption{\label{Fig10} $^1S_0$ - $^3P_1$ MOT temperature versus
(a) detuning $\delta$ for various intensities and (b) intensity
$s$ at $\delta$ = -520 kHz. In (a), DT is standard Doppler
theory. In (b), solid circles are experimental data while the
solid and dotted lines give the Doppler limit
($\hbar\Gamma_E$/2k$_B$) and the single-photon recoil limit
(2$\hbar\omega_R$/k$_B$), respectively. (c) Global scaling factor
applied to Doppler theory in order to match regime (II) data
versus intensity.}
\end{figure}

To obtain a quantitative expression for T$_M$ under these
conditions, we first find the damping coefficient $\alpha$ by
Taylor expanding
\begin{eqnarray}
F(v_z, z) &=& \frac{\hbar k\Gamma}{2}\left[\frac{s}{1 + s' +
\displaystyle\frac{4(\Delta - kv_z - \mu dB_z z)^2}{\Gamma^2}}\right]\nonumber \\
&-& mg. \label{Fe1}
\end{eqnarray}
where ($\partial F$/$\partial v$) = $\alpha$ is evaluated at
$v_z$ = 0, z = $z_0$. Solving $F(0,z_0)$ = 0 and using $R$ =
($\hbar k \Gamma$/$2mg$), the $\delta$-independent effective
detuning is given by
\begin{equation}
\frac{\Delta - \mu dB_z z}{\Gamma} = -\frac{\sqrt{Rs-s'-1}}{2}
\label{ED}
\end{equation}
which, in combination with Eq. \ref{Fe1}, states that the
scattering rate depends only on $R$, $s$, and $\Gamma$.
Substituting this expression into ($\partial F$/$\partial v$), we
obtain the damping coefficient
\begin{equation}
\alpha = -\frac{2\hbar k^2\sqrt{Rs-s'-1}}{R^2 s} \label{damp}
\end{equation}
Next, the diffusion coefficient $D_p$ is calculated by
substituting Eq. \ref{ED} into the single-beam scattering rate.
$D_p$ is then given by
\begin{equation}
D_p = \frac{\hbar^2 k^2 \Gamma}{2R} \label{diff}
\end{equation}
Combining Eqs. \ref{damp} and \ref{diff}, the predicted
$\delta$-independent equilibrium temperature is
\begin{eqnarray}
T = \frac{D_p}{|\alpha|} &=&
\frac{\hbar\Gamma\sqrt{s}}{2k_B}\frac{R}{2\sqrt{R-s'/s-1/s}}\nonumber
\\ &=& \left(\frac{\hbar\Gamma_E}{2k_B}\right)N_R \label{R1Temp}
\end{eqnarray}
where we have used $\sqrt{s}$ $\sim$ $\sqrt{s+1}$ for $s$ $\gg$
1. As shown by Fig. 11(b), the numerical factor $N_R$ is
approximately 2 over the entire relevant experimental range.
Regime (I) temperatures, therefore, depend only on $\Gamma_E$. To
test this prediction, Fig. 10(b) shows T$_M$ versus $s$ for a
fixed large detuning $\delta$ = -520 kHz. For the central portion
of the plot where regime (I) dynamics are relevant, we find good
agreement with Eq. \ref{R1Temp}.

\begin{figure}[t]
\resizebox{8cm}{!}{
\includegraphics{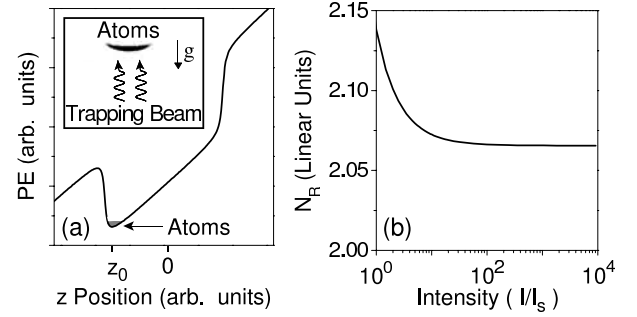}}
\caption{\label{Fig11} (a) Schematic depiction of regime (I)
cooling. Atoms sag to the bottom of the potential energy (PE)
curve where interactions occur primarily with the upward
traveling trapping beam. The inset shows the excitation geometry.
(b) Numerical factor N$_R$ in Eq. \ref{R1Temp} versus intensity.}
\end{figure}

For small $\delta$ in Fig. 10(a), the system evolves from regime
(I) to regime (II) where $\Delta$ $<$ $\Gamma_E$, $\Gamma_E$
$\gg$ $\Gamma$. As this transition occurs, trap dynamics change
from free-flight to damped harmonic motion. Here, one expects
thermodynamics similar to ordinary Doppler cooling including
$\delta$- and $s$-dependent minima with equilibrium values having
the functional form \cite{Lett, JILA1c}
\begin{eqnarray}
T(s,\Delta) & = & T_{0}(4|\Delta|/\Gamma_E)^{-1}\left(1 +
4(\Delta/\Gamma_E)^2\right) \nonumber\\
T_{0} & = & (\hbar\Gamma_E)/(2k_B) \label{SDT}
\end{eqnarray}
where $T_0$, realized at $\Delta$ = $\Gamma_E$/2, is a
generalized version of the $s$ $\ll$ 1 Doppler limit. As shown by
the solid lines in Fig. 10(a), Eq. \ref{SDT} correctly reproduces
the functional shape of the data. As shown by Fig. 10(c),
however, matching the absolute data values requires multiplying
Eq. \ref{SDT} by a $s$-dependent global scaling factor ($\leq$ 1)
whose value decreases with $s$, leading to temperatures well
below the standard Doppler limit $T_0$. In contrast to ordinary
Doppler cooling, the cloud thermal energy in regime (II) is thus
not limited by half the effective energy width of the cooling
transition. Notably, we find this surprising result cannot be
explained either by analytic treatments of Eq. \ref{SCF} or
semiclassical Monte-Carlo simulations of the cooling process. The
Monte-Carlo simulations, in fact, simply reproduce standard
Doppler theory. Finally, as $s$ approaches unity in Fig. 10(b),
the trap enters regime (III) where $\Gamma$ $\sim$ $\omega_R$
$\sim$ k$_B$T/$\hbar$ and thus the cooling becomes fully quantum
mechanical. Here, we obtain a minimum temperature of 250(20) nK,
in good agreement with the quantum mechanically predicted value
of half the photon recoil temperature $T_R$/2 =
$\hbar\omega_R$/k$_B$ = 230 nK \cite{dalibard89}.

\begin{figure}[t]
\resizebox{7.8cm}{!}{
\includegraphics{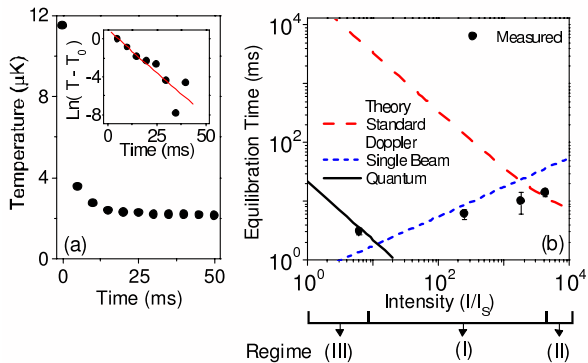}}
\caption{\label{Fig12} The approach to thermal equilibrium at
$\delta$ = -520 kHz. (a) Measured temperature versus time for $s$
= 248. The inset gives a semi-log plot for times greater than 5
ms. The solid line is a linear fit. T, instantaneous temperature;
T$_0$, final equilibrated temperature (b) Measured exponential
equilibration time versus $s$. The dashed, dotted, and solid
lines are predictions from standard Doppler theory, the
single-beam theory outlined in the text, and quantum theory,
respectively.}
\end{figure}

The approach to thermal equilibrium also displays signatures of
regimes (I) - (III). Figure 12(a) gives the cloud temperature
versus time for a $\delta$ = -520 kHz, $s$ = 248 single-frequency
MOT. As shown by the figure inset, the data is well fit by a
single exponential aside from the rapid decrease for times $<$ 5
ms that arises from atoms falling outside the trap capture
velocity \cite{Curtis, PTB}. Recall from Fig. 10 that scanning
$s$ at this detuning provides access to all three cooling
regimes, each of which is characterized by a unique evolution
toward thermal equilibrium. Specifically, for regimes (I) and
(II), semiclassical Doppler theory predicts and equilibration time
$\tau$ = $m$/2$|\alpha|$ where $\alpha$ is given in regime (I) by
Eq. \ref{damp} and, according to ordinary Doppler theory, in
regime (II) by \cite{Lett, JILA1c}
\begin{equation}
\alpha = -\frac{4 \hbar
k^2}{3}\frac{|\Delta|}{\Gamma}\left[\frac{s}{1 + s + 4(\Delta /
\Gamma)^2}\right] \label{SDalpha}
\end{equation}
Conversely, quantum theory predicts that for regime (III), $\tau$
follows \cite{dalibard89}
\begin{equation}
\tau \sim \frac{(k_B T_i)^2}{\hbar^2 \Gamma^3 s} \label{QMtau}
\end{equation}
where $T_i$ is the initial cloud temperature. As a test of these
predictions, Fig. 12(b) shows $\tau$ versus $s$ at $\delta$ =
-520 kHz along with $\tau$ values predicted by the above three
theories. Note that for each $s$ value we find the temperature
versus time is well fit by a single exponential aside from the
$\sim$ 5 ms long rapid decrease typified by Fig. 12(a). In regime
(II) at $s$ = 4520, $\tau$ = 14(2) ms, in good agreement with
Doppler theory which gives $\tau$ $\sim$ 12 ms. As expected for
regime (I), corresponding to 10 $<$ s $\leq$ 4000, $\tau$ follows
equilibration times predicted by the single-beam damping
coefficient while in regime (III) at $s$ = 6.1, $\tau$ = 3.1(4)
ms, consistent with the Eq. \ref{QMtau} quantum mechanical
predictions.

\section{$\delta$ $>$ 0 Cooling and Momentum-Space Crystals}

Tuning to $\delta$ $>$ 0 during the single-frequency cooling
stage reveals two fundamental and unique physical processes: (1)
the creation of well-defined momentum packets whose velocity
space alignment mimics lattice points on a face-centered cubic
crystal and (2) laser cooling around a velocity $v_0$ where the
radiative force balances gravity. In the following, we explore
these two effects in detail.

Basic insight into $\delta$ $>$ 0 momentum packet formation can
be obtained by considering the elementary problem of 1D atomic
motion in the presence of two counter-propagating $\delta$ $>$ 0
light fields. For simplicity, we assume $\vec{dB}$ = 0. According
to Eq. \ref{SCF}, an atom with initial velocity $\vec{v_i}$ will
preferentially interact with the field for which
$\vec{k}\cdot\vec{v_i}$ $>$ 0. Hence, the absorption process
preferentially accelerates rather than decelerates the atom,
further decreasing the probability for absorption events that
slow the atomic motion and enabling "positive feedback" in
velocity space. This process terminates for final velocities $v_f$
satisfying
\begin{equation}
v_f \gg \frac{1}{k}\left(\Delta +
\frac{\Gamma\sqrt{1+s}}{2}\right) \label{vfinal}
\end{equation}
which has a linear $\Delta$ dependence and scales with $\Gamma_E$.
Neighbor atoms with initial velocities around $\vec{v_i}$ undergo
similar "positive feedback" acceleration and ultimately achieve
final velocities near $v_f$. To quantify this latter effect, we
first simplify the problem by neglecting the beam for which
$\vec{k}\cdot\vec{v_i}$ $<$ 0. The equation of motion for the
atomic velocity is then
\begin{equation}
\frac{\partial v}{\partial t} = \frac{\hbar
k\Gamma}{2m}\left[\frac{s}{1 + s + \displaystyle\frac{4(\Delta -
kv)^2}{\Gamma^2}}\right] \label{ABlue}
\end{equation}
which can be solved analytically for the interaction time $t$ as
a function of the velocity $v$
\begin{eqnarray}
t(v,v_i) &=& \frac{2m}{\hbar k s \Gamma^3}[\frac{4}{3}k^2v^3 -
4\Delta kv^2 + ((1+s)\Gamma^2 + 4\Delta^2)v \nonumber\\ &-&
\frac{4}{3}k^2v_i^3 - 4\Delta kv_i^2 + ((1+s)\Gamma^2 +
4\Delta^2)v_i] \label{ASol}
\end{eqnarray}

\begin{figure}
\resizebox{8.2cm}{!}{
\includegraphics{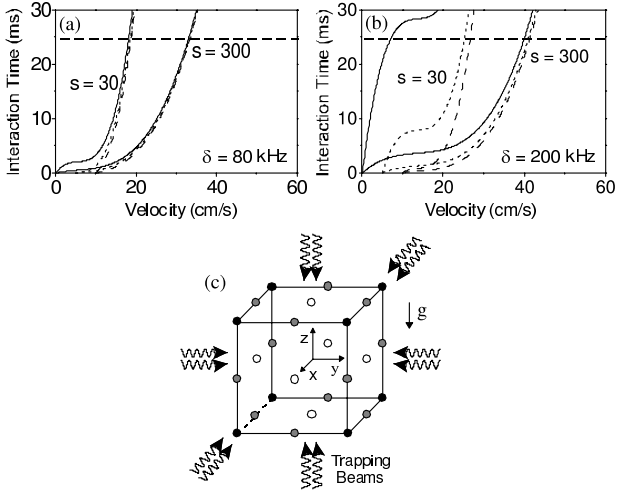}}
\caption{\label{Fig13} One dimensional velocity bunching due to
single-beam $\delta$ $>$ 0 absorption for (a) $\delta$ = 80 kHz,
and (b) $\delta$ = 200 kHz. Solid, short dash, and dashed lines
correspond to initial velocities of 0.1 cm/s, 5 cm/s, and 10
cm/s, respectively. (c) Three-dimensional $\delta$ $>$ 0 momentum
space structure. Final momenta corresponding to the black, gray,
and white circles arise from three-, two-, and single-beam
interactions, respectively.}
\end{figure}

\begin{figure}
\resizebox{8.5cm}{!}{
\includegraphics{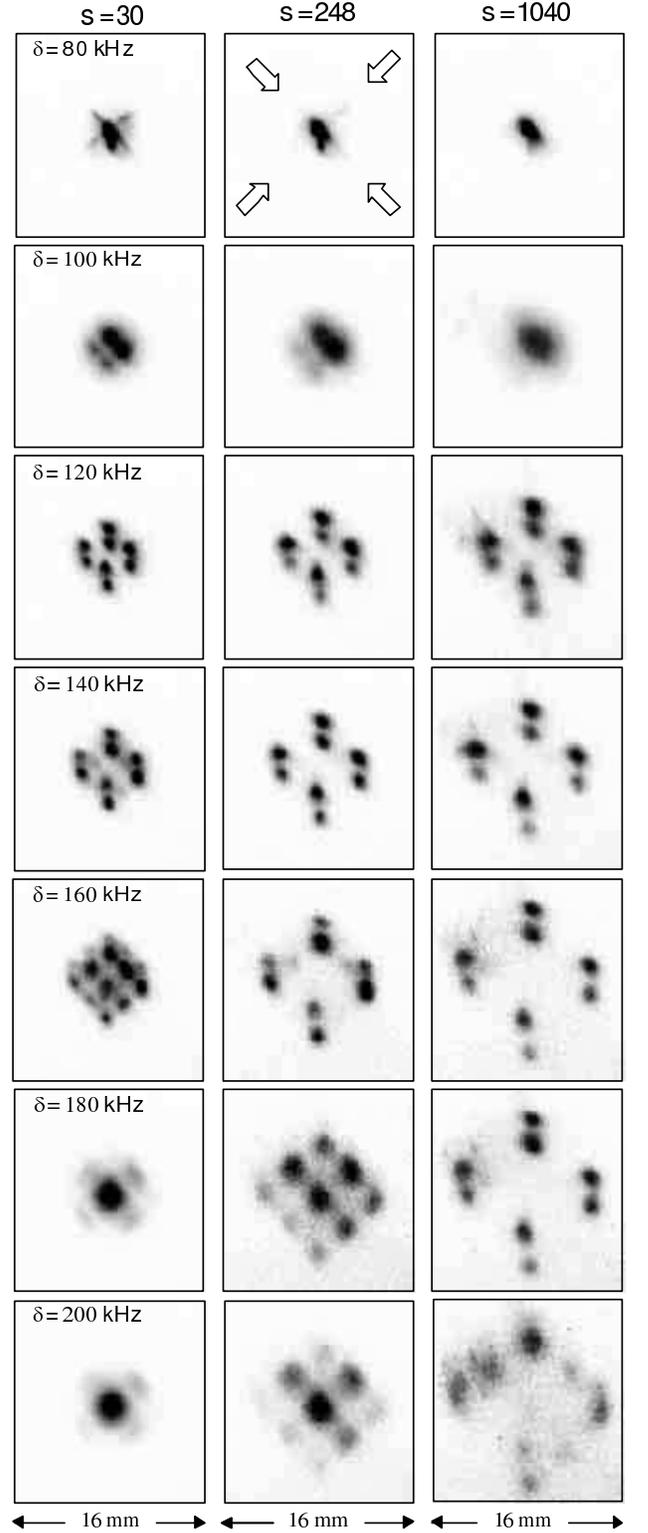}}
\caption{\label{Fig14} Top view in-situ images for a range of
$\delta$ and $s$ at $|\vec{dB}|$ = 0 and t$_H$ = t$_V$ = 25 ms.
Arrows in the uppermost center frame show the propagation
direction for the horizontal molasses beams. The initial cloud
temperature is 11.5(2) $\mu$K.}
\end{figure}

\noindent Figures 13(a) and 13(b) show plots of Eq. \ref{ASol} at
$\delta$ = 80 kHz and $\delta$ = 200 kHz, respectively. For both,
$s$ = 30 or $s$ = 300. The chosen $v_i$ values of 0.1 cm/s, 5
cm/s, and 10 cm/s span the velocity distribution for a 12 $\mu$K
cloud. Assuming an interaction time of 25 ms, Fig. 13(a) clearly
shows that due to laser induced acceleration, the entire range of
initial velocities is rapidly bunched to a significantly reduced
range of final velocities. Considering a fully 1D situation, this
result implies the cloud is divided into two well defined and
oppositely moving packets. As $\delta$ increases to 200 kHz in
Fig. 13(b), the $s$ = 30 atom-light interaction becomes
sufficiently weak for $v_i$ $\sim$ 0 that two final velocity
groups centered around $v_f$ $\sim$ 8 cm/s and $v_f$ $\sim$ 25
cm/s appear. In contrast, we find that for $s$ = 300 the
transition to two groups does not occur until $\delta$ $\sim$ 400
kHz. Accounting again for the full 1D symmetry, the $s$ = 30 case
corresponds to dividing the $t$ = 0 cloud into three groups, two
that move in opposite directions with relatively large final
velocities and one that, in comparison, is nearly stationary.
Overall, Eq. \ref{ABlue} thus predicts that for $\delta$ $>$ 0
the atomic cloud evolves into a discrete set of momentum-space
packets. As $s$ decreases at fixed $\delta$, the number of packets
increases while, as predicted  by Eq. \ref{vfinal}, the mean
velocity for a given packet scales with both $\delta$ and $s$.

Generalizing this analysis to the full 3D molasses beam geometry
leads to the structure shown in Fig. 13(c): a 3D array of
momentum-space groups whose alignment mimics the lattice points
on a face-centered-cubic crystal. From symmetry considerations,
cube corners correspond to three beam interactions in which the
cloud is divided into two pieces along each coordinate axis. From
Figs. 13(a) and 13(b), these points have a $\delta$- and
$s$-dependent mean velocity and appear at relatively small
detunings. For larger $\delta$ values, two-beam interactions fill
points along the Fig. 13(c) corner connecting lines. In analogy
to $s$ = 30 in Fig. 13(b), atoms in these points have $v_i$
$\sim$ 0 along a single axis and thus remain nearly stationary.
Along the remaining two axes, however, $v_i$ $>$ 0, enabling
acceleration to larger $v_f$. The 3-beam to 2-beam transition, as
shown in Figs. 13(a) and 13(b), occurs at progressively larger
$\delta$ values as $s$ increases. Together, the two processes
form a total of 20 divided groups with 8, 4, and 8 packets in the
top, middle, and bottom layers of the cube, respectively.
Finally, as $\delta$ increases further, atoms with $v_i$ $\sim$ 0
along two axes experience acceleration only along a single axis,
producing the 1-beam lattice points shown as 6 open circles in
the Fig. 13(c) cube face centers.

\begin{figure}
\resizebox{8.0cm}{!}{
\includegraphics{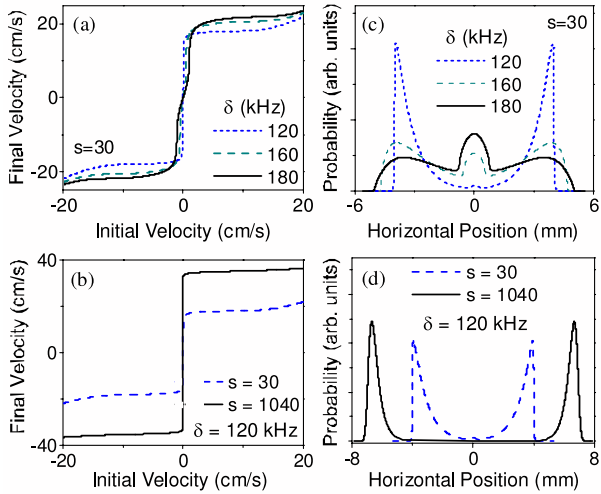}}
\caption{\label{Fig15} Calculated horizontal $\delta$ $>$ 0 final
versus initial velocity for (a) various $\delta$ at $s$ = 30 and
(b) various $s$ at $\delta$ = 120 kHz. (c) and (d) give the
corresponding spatial distributions. For each, t$_H$ = 25 ms and
the t$_H$ = 0 temperature is 11.5 $\mu$K.}
\end{figure}

Figure 14 shows an array of top view (slightly off vertical)
in-situ cloud images for intensities ranging from $s$ = 30 to $s$
= 1040 and detunings spanning $\delta$ = 80 kHz to $\delta$ = 200
kHz. For each, $|\vec{dB}|$ = 0 and the atom-light interaction
time is fixed at t$_H$ = 25 ms (t$_V$ = 25 ms) in the horizontal
x-y plane (along z-axis) molasses beams. The initial t$_H$ =
t$_V$ = 0 cloud temperature is 11.5(2) $\mu$K. Here, the observed
cloud evolution agrees well with the qualitative predictions
developed from the Fig. 13 model. As $\delta$ increases at fixed
$s$, sets of n-beam lattice points sequentially fill with n = 3
filling first followed by n = 2,1. Moreover, as $s$ increases,
transitions between the n-beam processes occur at progressively
larger $\delta$ values. Finally, for fixed $\delta$, the mean
lattice point velocity, proportional to the resulting lattice
point spacing in the Fig. 14 images, scales with $s$ as expected
from both Eq. \ref{vfinal} and Fig. 13. Note that qualitatively
similar dynamics are observed for $|\vec{dB}|$ $\neq$ 0. The
detailed changes to the cloud evolution that result from this
situation are described below.

\begin{figure}
\resizebox{8.5cm}{!}{
\includegraphics{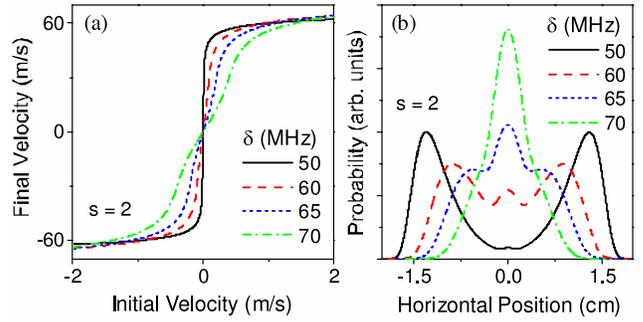}}
\caption{\label{Fig16} Broad line momentum space crystals.
Calculated horizontal (a) final versus initial velocity and (b)
corresponding spatial distributions for 461 nm $\delta$ $>$ 0
optical molasses. For each, t$_H$ = 500 $\mu$s and the initial
temperature is 2.5 mK.}
\end{figure}

The intuitive understanding of $\delta$ $>$ 0 dynamics developed
above is confirmed by Fig. 15. Here, we show numerical
calculations based on Eq. \ref{SCF} of the final horizontal
velocity and spatial distributions for the $s$ = 30 column and
$\delta$ = 120 kHz row in Fig. 14. For the calculations,
$|\vec{dB}|$ = 0, t$_H$ = 25 ms, and the initial cloud
temperature is 11.5 $\mu$K. The Figs. 15(c) and 15(d) spatial
distributions should be compared to cube lines in the x-y plane
along the Fig. 14 x-y molasses beam propagation directions.
Importantly, this fully 1D model reproduces both the Fig. 14
observations and the Fig. 13 predictions for the $\delta$- and
$s$-dependent lattice point filling factors and mean lattice point
velocity. As expected, the temperature of each packet in its
moving frame is lower than the t$_H$ = t$_V$ = 0 ms atomic cloud,
a result arising from the previously described velocity bunching
effects and directly connected to cloud shape asymmetries
observed in both the experiment (note the sharp outer cloud edges)
and the Fig. 15 theory. Notably, however, only two vertical
layers are observed in Fig. 14 while the Fig. 13 model predicts
three. As explained below, this apparent discrepancy arises from
novel gravitationally induced z-axis dynamics.

At this point, it is important to realize that Eq. \ref{SCF} is
semiclassical and thus does not account for dynamics that depend
on the relative size of $\Gamma$ and $\omega_R$. The $\delta$ $>$
0 momentum crystals observed here, therefore, are a universal
feature of Doppler limited systems. Similar processes should then
occur with broad lines such as the 461 nm $^1S_0$ - $^1P_1$
transition. To test this possibility, Fig. 16 shows numerically
simulated final horizontal velocity and spatial distributions for
a 2.5 mK $^{88}$Sr cloud excited by 461 nm $\delta$ $>$ 0 optical
molasses. For the calculation, $|\vec{dB}|$ = 0, t$_H$ = 500
$\mu$s, and $s$ = 2. As clearly indicated by the figure,
structures very similar to those shown in Fig. 14 can be
generated over length scales consistent with 461 nm cooling beam
diameters and typical $^1S_0$ - $^1P_1$ MOT temperatures.
Moreover, we find that by increasing either $s$ or t$_H$,
structures with contrast identical to those shown in Figs. 14 and
15 can be created. Note the model does not account for
spontaneous emission induced random walk heating, a process the
would tend to smear the contrast between individual momentum
packets. This omission, however, should not significantly affect
Fig. 16 since random walk heating scales as $\sqrt{\Gamma}$
\cite{Lett} while, from Eq. \ref{SCF}, the directional
acceleration scales as $\Gamma$. Thus, as $\Gamma$ increases from
the narrow to the broad line case, random walk heating becomes
progressively less important.

\begin{figure}
\resizebox{7.8cm}{!}{
\includegraphics{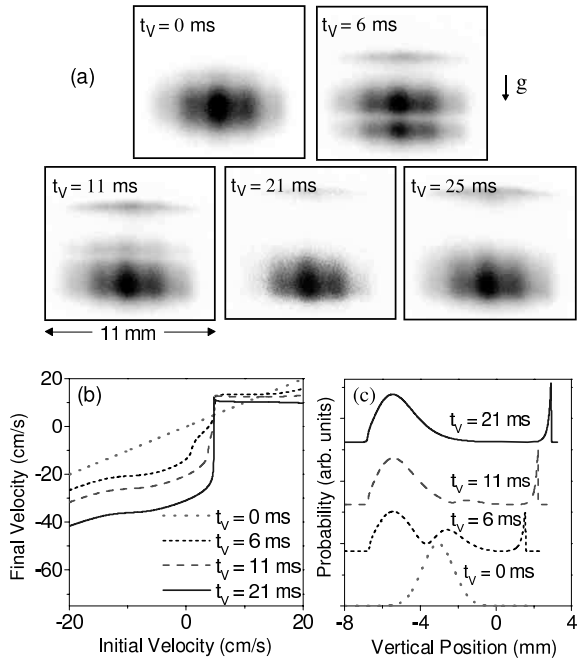}}
\caption{\label{Fig17} Side view in-situ images for $s$ = 30,
$|\vec{dB}|$ = 0, $\delta$ = 140 kHz, t$_H$ = 25 ms, and a range
of t$_V$ times. The t$_H$ = t$_V$ = 0 cloud temperature is 11.5(2)
$\mu$K. Calculated (b) final versus initial velocity for the same
set of parameters. (c) Corresponding calculated spatial
distributions.}
\end{figure}

Although the physics underlying $\delta$ $>$ 0 momentum-space
crystals is thus fully operative for broad lines, details of the
crystal formation process will make experimental observations
difficult. From Eq. \ref{vfinal}, $v_f$ for a given packet is
proportional to $\Gamma$. Hence, broad line packet velocities are
orders of magnitude larger than those achieved in the narrow line
case. In Fig. 16, for example, $v_f$ $\sim$ $\pm$ 60 m/s.
Visualizing the entire structure then requires imaging light with
an optical bandwidth of $\sim$ 260 MHz or an equivalently broad
optical transition. This situation should be compared to the
experiments performed here where $v_f$ $<$ 1 m/s, and hence
sufficient imaging bandwidth is obtained with the $\Gamma$/2$\pi$
= 32 MHz $^1S_0$ - $^1P_1$ transition. In addition, $\Gamma$ sets
the lattice point spacing, or alternatively the molasses beam
diameter, required to achieve a given contrast between individual
lattice points. For the contrast shown in Figs. 14 and 15, for
example, the required broad line molasses beam diameters grow to
tens of centimeters, making experimental observations impractical.

Figure 17(a) shows in-situ images of the 689 nm $\delta$ $>$ 0
molasses when the cloud is viewed in the horizontal x-y plane at
45$^o$ to the x-y axes. For the images, $|\vec{dB}|$ = 0, $s$ =
30, $\delta$ = 140 kHz, t$_H$ = 25 ms, and t$_V$ is varied from
t$_V$ = 0 ms to t$_V$ = 25 ms. Here, the apparent contradiction
between Fig. 14 and the Fig. 13(c) prediction for the number of
vertical layers is shown to occur due to gravity induced
dynamics. Increasing t$_V$ from t$_V$ = 0 ms to t$_V$ = 6 ms, for
example, creates three vertical layers, as predicted by Fig.
13(c). For t$_V$ $>$ 6 ms, however, the lower two layers slowly
merge together, becoming a single cloud along the vertical
direction for t$_V$ $>$ 21 ms. To understand this process, recall
that the central layer in the Fig. 13(c) cube corresponds to
atoms with near zero z-axis velocities. For the chosen $\delta$
and in the absence of gravity, therefore, these atoms would
remain near $v_z$ = 0. As shown by the numerical calculations in
Figs. 17(b) and 17(c), however, gravity accelerates the central
layer into resonance with the downward propagating molasses beam,
causing the two downward propagating layers to merge. For the
t$_V$ = 25 ms time used in Fig. 14, this process is complete.
Hence, only two layers are observed with the more (less) intense
packets in the Fig. 14 images corresponding to the lower two
(uppermost) cube layers.

\begin{figure}
\resizebox{7.0cm}{!}{
\includegraphics{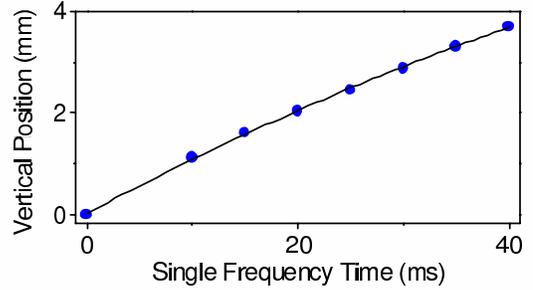}}
\caption{\label{Fig18} Measured vertical position of the upward
moving layer versus t$_V$ at $\delta$ = 140 kHz and $s$ = 30. The
solid line is a fit to the model described in the text.}
\end{figure}

Comparing the numerically simulated $\delta$ $>$ 0 spatial
distributions with Figs. 14 and 17(a) reveals that theoretically
predicted $|\vec{dB}|$ = 0 lattice spacings are $\sim$ 2$\times$
larger than observed. This result occurs due to stray magnetic
fields. Here, an independently measured $\sim$ 100 mG/cm
permanent chamber magnetization spatially shifts the effective
detuning as the atoms move outward, causing an apparent
deceleration and thus reduced lattice spacing. This effect can be
seen most clearly by measuring the vertical position of the
upward moving cube layer $z_T$ versus t$_V$. For the upward
moving layer, Eq. \ref{SCF} predicts that the atoms are
accelerated to a velocity $v_0$ where the radiative force
balances gravity. The cloud then moves upward at $v_0$ and hence,
experiences an effective acceleration $g_e$ = 0. To test for a
magnetic field induced non-zero $g_e$, Fig. 18 shows $z_T$ versus
t$_V$ for $s$ = 30 and $\delta$ = 140 kHz. The solid line in the
figure is a fit to the simple kinematic equation $z_T$ = $z_0 +
v_0 t_V$ - $(g_e t_V^2)/2$. Here, $v_0$ is treated as an initial
velocity since for $s$ = 30 and $\delta$ = 140 kHz, the atoms are
accelerated to $v_0$ in $<$ 2 ms. From the fit, we find $g_e$ =
-0.98(16) m/s$^2$, consistent with a stray gradient dB$_z$ = 100
mG/cm. Once this gradient is included in numerical calculations
of Eq. \ref{SCF}, the measured and calculated lattice point
spacings agree.

Finally, the exceptionally sharp velocity and thus spatial
distributions for the upward propagating layer in Fig. 17 imply
velocity compression beyond the bunching effects described
earlier. In fact, as shown by Fig. 19(a) which depicts the
composite gravitational and radiative force for the upward moving
atoms, stable cooling occurs around $v_0$ where the composite
force is zero. From Eq. \ref{Fe1}, $v_0$ is given by
\begin{equation}
v_0 = \frac{1}{k}\left(\Delta +
\frac{\Gamma}{2}\sqrt{s(R-1)-1}\right) \label{v0Eq}
\end{equation}
which depends linearly on $\Delta$ and scales approximately as
$\Gamma_E$. Fig. 19(b) shows measured values for $v_0$ versus
$\delta$ at $s$ = 10 and $s$ = 75, clearly demonstrating the
expected linear $\delta$-dependence. To obtain quantitative
comparisons with theory, we next perform linear fits to the data
and then obtain predicted values for $v_0$ from numerical
calculations that include the 100 mG/cm field gradient discussed
above. From the experiment, we find ($\partial v_0$/$\partial
\delta$) = 644(17) $\mu$m/(kHz-s) at s = 10 and 634(19)
$\mu$m/(kHz-s) at s = 75. These slopes agree at the 10$\%$ level
with the predicted values of ($\partial v_0$/$\partial \delta$) =
572(5) $\mu$m/(kHz-s) at s = 10 and 638(20) $\mu$m/(kHz-s) at s =
75. Moreover, experimentally observed absolute values for $v_0$
are reproduced by the calculations at the level of 20$\%$, in
good agreement with the expected Eq. \ref{v0Eq} intensity
dependence.

\begin{figure}
\resizebox{8.5cm}{!}{
\includegraphics{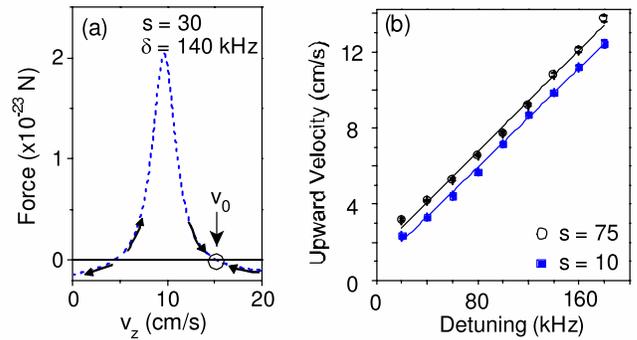}}
\caption{\label{Fig19} $\delta$ $>$ 0 cooling. (a) Composite
gravitational and $\delta$ $>$ 0 radiative force versus v$_z$.
Arrows show the direction for force induced velocity changes.
Stable cooling occurs around a velocity $v_0$, labeled by a
circle, where gravity cancels the radiative force. (b) Measured
$v_0$ versus $\delta$ for $s$ = 10 and $s$ = 75. Solid lines are
linear fits.}
\end{figure}

The equilibrium temperature of the upward moving layer can be
calculated by following the same procedure used to derive Eq.
\ref{R1Temp}. Here, however, we set $\mu dB_z z$ = 0 in Eq.
\ref{Fe1} and perform the Taylor expansion about $v_z$ = $v_0$.
The $\delta$-independent effective detuning is then
\begin{equation}
\frac{\Delta - k v_z}{\Gamma} = -\frac{\sqrt{Rs-s'-1}}{2}
\label{ED2}
\end{equation}
which leads to damping and diffusion coefficients identical to
Eq. \ref{damp} and \ref{diff}, respectively. Hence, the expected
equilibrium temperature is given by Eq. \ref{R1Temp}. Thus,
gravity, via the ratio $R$, again plays an important role in
narrow line thermodynamics, in this case enabling $\delta$ $>$ 0
cooling. Unfortunately, this prediction cannot be accurately
verified due to spatial overlap among the horizontal plane
packets in the upward moving cube layer. The observed and
theoretically predicted sharp vertical spatial distributions in
Figs. 17(a) and 17(c), however, strongly suggest that $\delta$
$>$ 0 cooling is operative in the experiment. Finally, note that
although this same cooling mechanism should occur for broad lines
where $R$ $\sim$ 10$^5$, $v_0$ in this case has impractical
values on the order of 100 m/s. Moreover equilibrium temperatures
are large, at roughly 160$\times$($\hbar \Gamma_E$/2$k_B$) $\sim$
200 mK for the $^1S_0$ - $^1P_1$ transition even at $s$ = 1.

\section{Conclusions}

In summary, narrow line laser cooling exhibits a wealth of
behaviors ranging from novel semiclassical dynamics wherein
gravity can play an essential role to quantum mechanically
dominated sub-photon recoil cooling. In the $\delta$ $<$ 0
semiclassical case, trap dynamics are set by either hard wall
boundaries or a linear restoring force. Qualitative differences
between these two situations are reflected in both the atomic
motion and the equilibrium thermodynamics. Here, mechanical
dynamics range from free-flight in a box potential to damped
harmonic oscillation. Accordingly, equilibrium temperatures range
from detuning independent values scaled by the power-broadened
transition linewidth to detuning dependent minima well below the
standard Doppler limit. As the saturation parameter approaches
unity, the trap enters a quantum mechanical regime where
temperatures fall below the photon recoil limit despite the
incoherent trapping beam excitation. For $\delta$ $>$ 0, the
cloud divides into momentum-space crystals containing up to 26
well defined lattice points and the system exhibits $\delta$ $>$ 0
gravitationally assisted cooling. These surprising $\delta$ $>$ 0
behaviors, which again occur due to an incoherent process, are
theoretically universal features of Doppler limited systems.
Observations should therefore be possible, although difficult,
with broad line optical molasses. Perhaps a similar $\delta$ $>$
0 mechanical evolution also occurs for atoms, such as the more
typically employed Alkali metals, that support both Doppler and
sub-Doppler cooling.

The authors wish to thank K. Hollman and Dr. R. J. Jones for their
work on the femto-second comb measurements. This work is funded by
ONR, NSF, NASA, and NIST.


\begin{thebibliography}{10}
\bibitem{Com1}
    This analogy is not rigorously correct since
    face-centered-cubic crystals lack the lattice points along
    corner-connecting lines that are observed in the experiment.
\end{thebibliography}
\end{document}